# A dynamic backrest may help accommodate trunk movement, and multi-device work elicits more of it: An exploratory comparison of high-end office chairs under single- and multi-device tasks

Yejin Kang[a, b], Dusan Baek[b], Dongwoo Park[b], Kitae Hwang[b], Woojin Park[b, c, *]

[a] Industrial and Operations Engineering, University of Michigan, Ann Arbor, United States

[b] Department of Industrial Engineering, Seoul National University, Seoul, Republic of Korea

[c] Institute for Industrial Systems Innovation, Seoul National University

[*] Corresponding author.

Department of Industrial Engineering, Seoul National University, 1 Gwanak-ro, Gwanak-gu, Seoul, 08826, South Korea

Tel: +82 (02) 880 4310

E-mail address: woojinpark@snu.ac.kr

**Abstract:**

Dynamic-sitting office chairs are usually compared with static sitting rather than with each other, and the contribution of backrest deformability (lateral backrest movement) is unknown. Office work also increasingly spans multiple devices. Twenty healthy adults (10 men, 10 women) completed a 15-minute single-device and a 15-minute multi-device proofreading task on each of three chairs: two commercial dynamic-sitting chairs (T90, Aeron) and a T90 whose backrest lateral movement was mechanically fixed (T90-fixed). We recorded subjective ratings, head and trunk posture (inertial measurement units), upper trapezius and lateral gastrocnemius activity (electromyography), and task performance. Chair effects were largely null: T90 and Aeron did not differ in ratings or performance, posture and muscle activity showed only isolated, direction-inconsistent differences, and preferences were polarized. The standard T90 showed greater trunk movement (mean absolute angular velocity in yaw) than T90-fixed during the multi-device task ($p = 0.042$), which may indicate that its movable backrest helps accommodate trunk movement under demanding work; no other whole-sample posture, muscle, rating or performance measure differed between these two configurations, and replication is needed. Task effects were large and consistent: the multi-device task elicited greater and faster head, trunk and shoulder movement and higher upper trapezius activity than the single-device task, with lower reading throughput. Multi-device work thus imposes substantially greater movement demands than single-device work, which supports designing chairs that accommodate movement in modern work environments.

# 1. Introduction

Office ergonomics research has traditionally treated seated computer work as centered on a single screen and a single point of visual and manual attention. That assumption no longer matches how many people actually work. Interacting with more than one digital device at a time, switching between a laptop, an external monitor, a tablet, or a phone, has become a routine part of daily information work rather than an exception (Yuan et al., 2022), and multi-monitor setups are now common precisely because they offer more usable screen area (Gallagher et al., 2021). Layered onto this shift is the older, still unresolved problem of prolonged sitting itself: extended, largely motionless computer work continues to be linked to musculoskeletal discomfort and fatigue that accumulate over the course of a workday. One design response to this problem has been dynamic sitting, chairs engineered to allow, or actively invite, small in-seat postural changes such as shifting body weight, tilting the pelvis, or letting the backrest respond to trunk movement, rather than holding the sitter in one fixed configuration.

The term dynamic sitting denotes the additional postural motion that a chair or piece of equipment makes possible during seated work, as distinct from sitting understood as a fixed position (O'Sullivan et al., 2013). The rationale for building this capacity into chairs rests on two related lines of evidence. First, motion of this kind is understood to lower discomfort by keeping the trunk from settling into one sustained posture, which is why chairs and devices designed around it have been put forward as a way to offset the harms of prolonged, static sitting (Al-Eisa et al., 2013; O'Sullivan et al., 2012; O'Sullivan et al., 2006; Pynt, 2015). In line with this, Nüesch et al. (2018) showed that a commercial dynamic chair (Haider BioSwing) generated significantly more seat motion than an otherwise identical chair with its swinging system locked, with total sway path length clearly elevated in the dynamic condition ($p < 0.05$). Second, the logic extends specifically to low back pain. People with low back pain tend not to move subtly while seated but instead hold prolonged end-range postures punctuated by occasional large shifts, in contrast to the small, continual spinal adjustments seen in pain-free sitters, an observation that has motivated dynamic sitting as a candidate strategy for reducing low back pain during seated work (van Dieën et al., 2001).

The equipment used to study these effects on trunk muscle activation has varied widely in form. Several studies removed the backrest altogether, relying on exercise balls (Gregory et al., 2006; Kingma & van Dieën, 2009; McGill et al., 2006) or an air-filled cushion set on a stool (O'Sullivan et al., 2006). Others kept the backrest and built the dynamic element into it or around it: a continuous passive motion (CPM) device that periodically shifted lumbar curvature (Beach et al., 2003), four backed chairs each allowing a different pattern and degree of movement (Ellegast et al., 2012), and two backed chair variants with their own moving parts (van Dieën et al., 2001). Across this varied equipment, however, the findings do not consistently favor dynamic sitting: a systematic review found no difference in trunk muscle activation in five of the seven studies that compared dynamic with static sitting (O'Sullivan et al., 2013), and a dynamic chair increased chair sway but only selectively altered trunk muscle activation (Nüesch et al., 2018).

This body of evidence, however, largely predates the current generation of high-end office chairs, and it points to a first gap: although a few studies have compared several dynamic chairs with a conventional chair (Ellegast et al., 2012), the literature has mostly benchmarked dynamic chairs against static sitting, rather than pitting specific contemporary products directly against each other. This gap matters because chair interface design does not stand still: the mechanisms by which adjustability and movement are engineered change substantially across product generations, so results obtained from older or discontinued models offer limited guidance about the chairs sold and used today. Establishing whether today's dynamic chairs actually deliver on their design intent therefore requires comparing specific, currently sold products directly against one another, not only against a static baseline.

The second gap follows directly from the equipment differences noted above. Because some of the dynamic chairs studied to date had no backrest (O'Sullivan et al., 2006) and others built their dynamic mechanism around one (Beach et al., 2003; Ellegast et al., 2012; van Dieën et al., 2001), the existing literature cannot say whether the backrest itself, and in particular its ability to flex or move in response to the trunk, contributes meaningfully to the benefits associated with dynamic sitting, or whether those benefits arise mainly from other parts of the chair. Answering this question requires isolating the backrest as a single design feature while holding every other aspect of the chair constant. Prior work has isolated other dynamic mechanisms in this way, for example by comparing an otherwise identical chair with its swinging system unlocked or locked (Nüesch et al., 2018), but, to our knowledge, not the backrest.

The third gap concerns the work itself rather than the chair. Nearly all of the dynamic-sitting evidence reviewed above comes from single-screen, single-device tasks, yet, as noted above, contemporary office work increasingly spans multiple devices used concurrently (Gallagher et al., 2021; Yuan et al., 2022). Multi-device work differs from single-screen work in ways that plausibly change postural and muscular demand: it requires dividing visual attention and manual interaction across separate devices and, when those devices are physically separated, reorienting the trunk and head to face each one in turn. Whether these added demands translate into measurably different posture, muscle activity, or task performance compared to conventional single-screen work has not been tested, yet it is a precondition for knowing whether chairs designed and evaluated under single-screen assumptions remain adequate as multi-device work becomes standard.

We designed the present study to close these three gaps at once, using three chairs and two tasks. The first two chair conditions are commercial products chosen specifically because they diverge in how they pursue dynamic sitting: the T90 produced by Sidiz, Pyeongtaek, Republic of Korea, and the Aeron produced by Herman Miller, Zeeland, Michigan, USA. Both sit at a comparable price point and level of design ambition, and both manufacturers build in-seat movement into their design rationale, yet each achieves this through its own engineering approach, with the two designs differing not only in the backrest but across most of their adjustable mechanisms. Comparing chairs that are similar in ambition but different in execution makes it possible to ask not just which one performs better, but what about its design explains the difference. The third condition isolates one of those design differences directly: a T90 prototype with its dynamic backrest mechanism mechanically locked in a fixed position (T90-fixed), otherwise identical to the standard T90.

Under each of the three chair conditions, participants completed two office tasks that differed in their device demands: a single-device task representative of conventional single-screen work, and a multi-device task that required working across two monitors and a smartphone and periodically turning away from and back to the workstation. Pairing these two tasks makes it possible to test directly whether the subjective ratings, postural, muscular, and performance demands of multi-device work differ from those of conventional single-screen work, independently of which chair is used.

The present study therefore pursues three specific objectives. The first is to determine whether two contemporary dynamic-sitting chairs that reach that goal through markedly different mechanisms, the T90 and the Aeron, actually differ in the subjective ratings along with postural, muscular, and performance outcomes they produce, thereby testing whether commercial dynamic chairs currently on the market deliver comparable benefits or diverge depending on how they are engineered. The second is to isolate the specific contribution of backrest deformability to any such benefits by comparing the standard T90 against the otherwise identical T90-fixed, allowing us to attribute observed effects to the backrest mechanism itself rather than to the chair's broader design. The third is to determine whether subjective ratings, postural, muscular, and performance outcomes differ between conventional single-device work and multi-device work, testing whether the added demands of dividing attention and interaction across multiple devices are measurable relative to single-screen work. Throughout both tasks we recorded subjective ratings, motion with inertial measurement units (IMUs), muscle activity with electromyography (EMG), and task performance, allowing us to answer all three questions with objective measures alongside self-report.

Findings from this study carry three distinct practical implications. First, by testing whether the T90 and the Aeron actually differ in the benefits they produce, they clarify whether commercially available dynamic chairs deliver comparable outcomes or diverge depending on how they are engineered, informing how dynamic office chairs should be selected for work that increasingly spans several devices at once. Second, by isolating the T90's backrest through the T90-fixed comparison, they show whether backrest deformability specifically drives these benefits, pointing to which design feature manufacturers should prioritize when engineering future dynamic chairs. Third, by comparing single-device and multi-device work directly, they indicate whether chair designs and evaluation standards developed around conventional single-screen work remain adequate as multi-device work becomes the norm, or whether office seating needs to be reconsidered for this emerging way of working.

## 2. Methods

### 2.1. Subjects

The study sample consisted of twenty healthy adults (10 men, 10 women) with no history of neck or back pain. To keep participants' evaluations of the chairs free from brand-based bias, we screened for and excluded anyone already familiar with the Herman Miller or Sidiz brands.

Men averaged 25.6 years of age (SD = 2.4), while women averaged 35.5 years (SD = 11.6). Standing height averaged 1.77 m (SD = 0.07 m; range: 1.68–1.86 m) among men and 1.62 m (SD = 0.06 m; range: 1.56–1.74 m) among women, and body weight averaged 70.4 kg (SD = 12.2 kg; range: 53–88 kg) for men versus 56.5 kg (SD = 4.5 kg; range: 50–65 kg) for women.

**2.2. Office chairs**

We selected two commercially available models a priori as theoretically informative contrasts: the T90, manufactured by the Korean company named Sidiz, and the Aeron, manufactured by the American company named Herman Miller. Both chairs reflect considerable ergonomic designs, and both manufacturers foreground in-seat movement and postural change as part of their design philosophy, though they articulate this goal through different design language. Despite this shared emphasis, the two chairs achieve it through substantially different mechanisms and overall chair architectures, which is precisely what makes them useful for isolating which design solutions are more effective and why.

To probe one specific source of that architectural difference, namely the backrest, we added a third condition consisting of a modified T90 prototype in which the backrest's lateral (left–right) movement was mechanically fixed in place. Backrest deformability refers to the ability of the backrest to move laterally in response to the sitter's trunk movement, and fixing it disabled this ability while leaving every other component of the chair unchanged (namely T90-fixed). Comparing this fixed-backrest prototype against the standard T90 isolates the specific contribution of backrest deformability, independent of the many other design differences that distinguish the T90 from the Aeron. Throughout the experiment, the chairs' brand identities were concealed from participants: the T90 was presented as Chair A, the T90-fixed prototype as Chair B, and the Aeron as Chair C. As Chair C lacked a headrest in its original configuration, a widely used third-party headrest of mesh-only construction (also sold together with the Aeron by authorized dealers) was added to allow comparison of this function with Chair A. This headrest differs in construction from that of the T90 (Chair A), which combines double raschel mesh and sponge. Front view of the three chairs are shown in Figure 1, whereas rear view of the three chairs are shown in Figure 2.

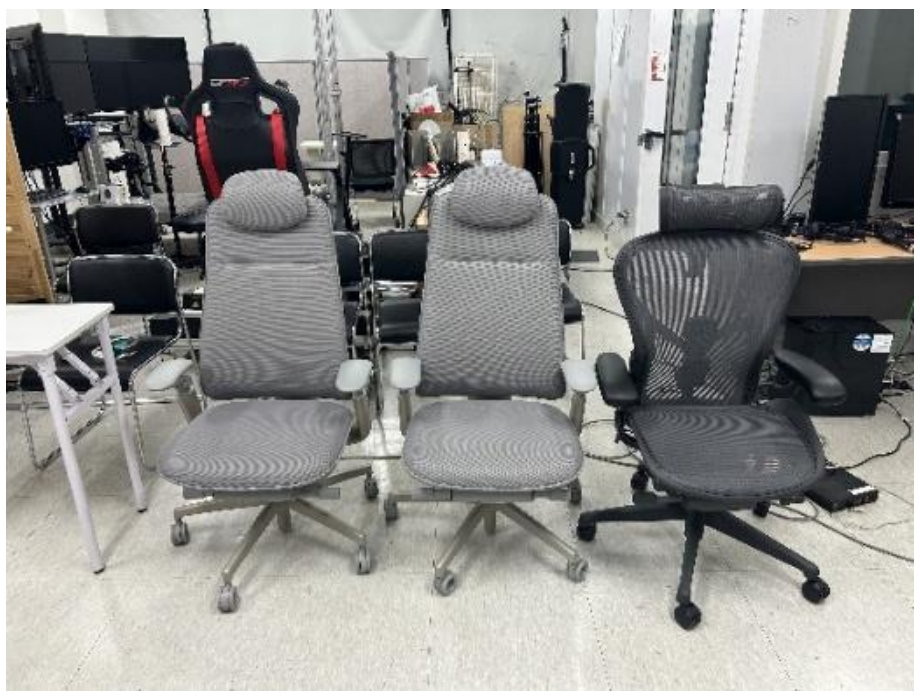

Figure 1. (From the left, front view) T90 (Chair A), T90-fixed (Chair B) and Aeron (Chair C)

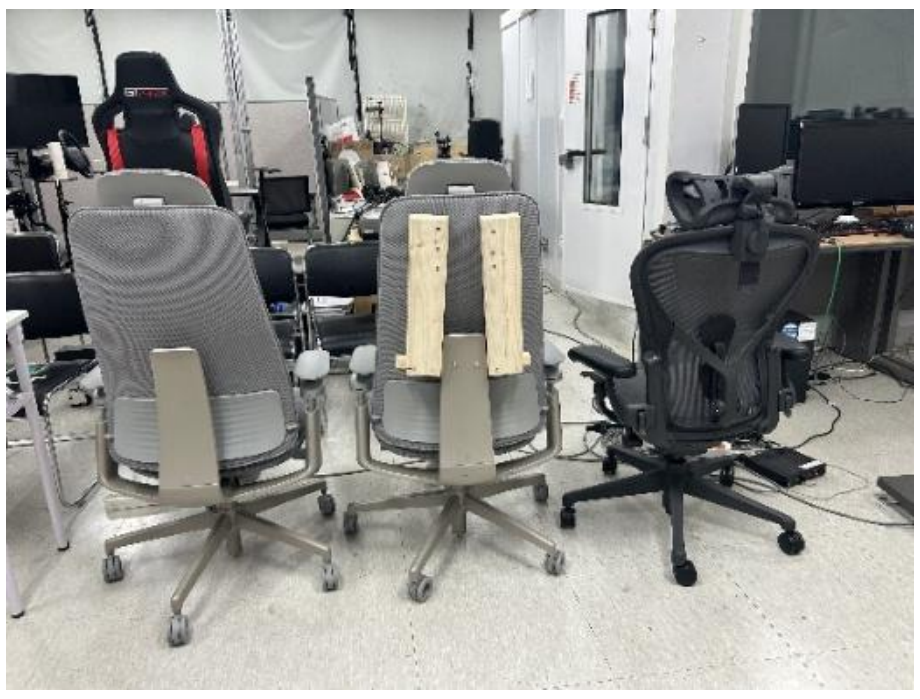

Figure 2. (From the left, rear view) T90 (Chair A), T90-fixed (Chair B) and Aeron (Chair C)

### 2.3. Task

Chair order and task order were counterbalanced across participants. Under each chair condition, participants completed two 15 minutes office tasks, one modeled on conventional single-device work and one modeled on the multi-device configurations. Each task was followed by subjective ratings, while EMG data, IMU data, task performance were recorded continuously throughout tasks.

#### 2.3.1. Task 1: Single-device task

In the single-device task (namely task 1), participants worked at one computer displaying two windows side by side on a single monitor. One window showed the original text of a novel excerpt and the other showed a modified version containing a series of wording changes. Participants compared the two windows sentence by sentence, identified the altered words, and corrected the modified text to match the original. This comparison and correction cycle continued for the full 15 minutes of the task. The setup is shown in Figure 3.

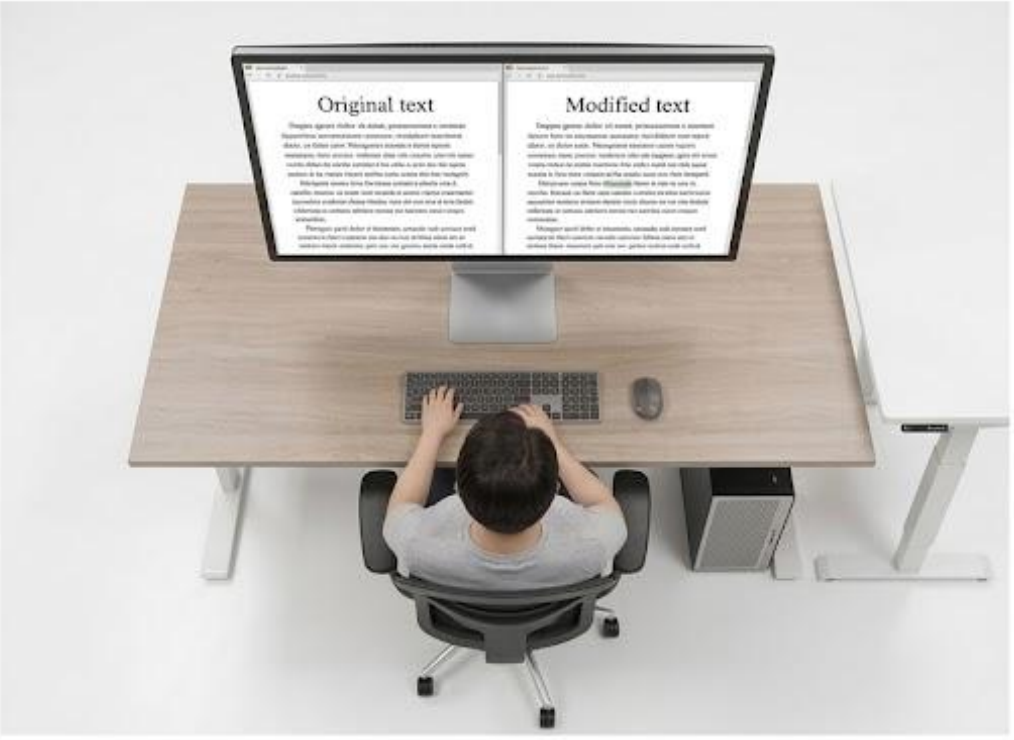

Figure 3. Experimental setup for task 1

#### 2.3.2. Task 2: Multi-device task

The multi-device task (namely task 2) used the same underlying comparison and correction procedure but distributed it across two computers and a smartphone. At the start of the task, participants retrieved a smartphone that had been charging at the corner of the desk. The original and modified text were then presented one window per computer, on two separate monitors, and participants compared and corrected the text in the same manner as in task 1. Each time a word was corrected, participants entered that word into the smartphone. In addition, at 3-minute intervals participants rotated 180 degrees to face a small human sculpture positioned behind them, verbally reported the words they had corrected since the previous report, and then returned to their original orientation to resume the task. This cycle of comparison, correction and periodic rotation continued for the full 15 minutes of the task. The setup is shown in Figure 4.

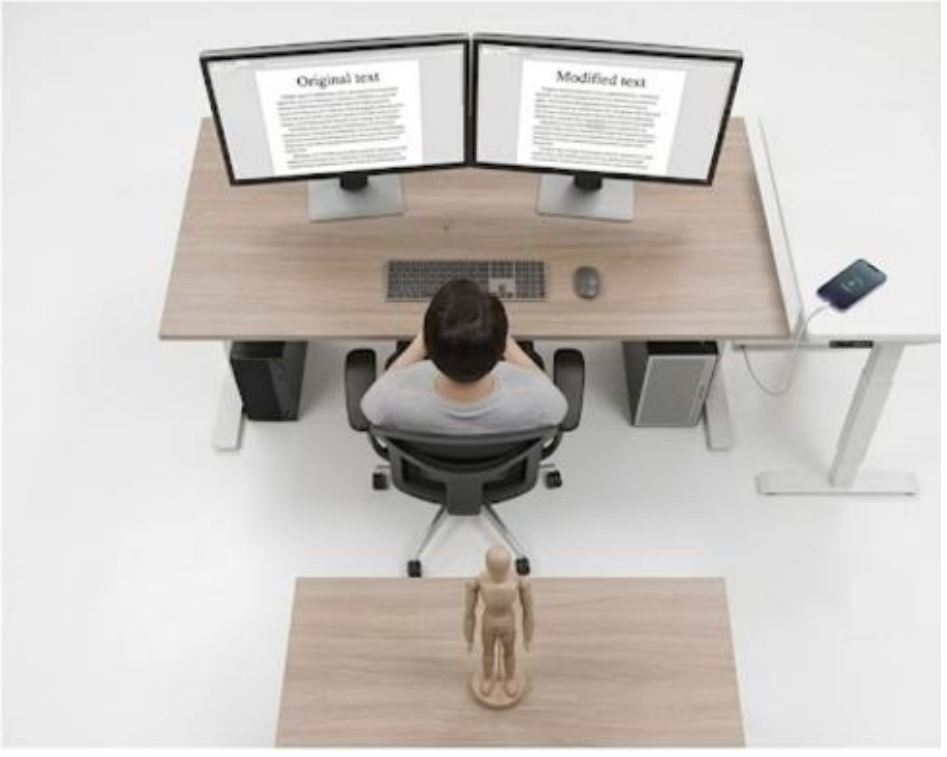

Figure 4. Experimental setup for task 2

### 2.4. Measurements

#### 2.4.1. Subjective ratings

After each task, participants rated their overall satisfaction with the chair and the seamless support it provided. Overall satisfaction was rated on a 5-point scale (1 = very dissatisfied to 5 = very satisfied). Seamless support, defined as the extent to which the chair continued to support the body as posture and movement changed during the task, was rated on the same 5-point scale (1 = poor to 5 = excellent).

After completing both tasks on all three chairs, participants freely ranked the three chairs (Chair A, Chair B, Chair C) from most to least preferred. Ties were permitted and given equal rank. Participants were then interviewed about the reasons for their ranking.

#### 2.4.2. IMU

Inertial measurement units (IMUs; Noraxon, Ultium, USA) were used to continuously record the orientation and position of the head and upper trunk at a sampling rate of 400 Hz during each task. Sensors were placed on the head, upper trunk, and left and right shoulders. Angular motion was described in terms of pitch (flexion–extension, i.e., forward–backward tilt), roll (lateral bending, i.e., left–right tilt), and yaw (axial rotation, i.e., left–right twist). Neck angle was defined as the orientation of the head segment relative to the trunk segment, approximating the joint angle at the seventh cervical vertebra (C7). Trunk angle was defined as the orientation of the upper-trunk segment relative to the vertical. Before the tasks, each participant's neutral upright posture was recorded and used as the zero-angle reference for angle computations. IMU data from one participant (a female) were excluded because of a technical failure during data acquisition. Accordingly, the IMU analyses were based on data from the remaining 19 participants. Raw angular signals were low-pass filtered with a fourth-order Butterworth filter at a cutoff frequency of 6 Hz. Signal components above 6 Hz during trunk movement were considered to reflect measurement noise rather than human movement (Brouwer et al., 2021), whereas typical activities of daily living fall largely within the 0.5–3.5 Hz range (Plasqui et al., 2013; Sun & Hill, 1993).

The analysis window extended from 0 to 900 s for each task, matching the window used for the EMG analysis (Section 2.4.3). In task 2, IMU signals were used to identify periods during which participants rotated 180° to report the corrected words and then returned to the front-facing posture. These periods were excluded for the same reason described for the EMG analysis; the task 2 results therefore represent the front-facing portion of the task. Five posture measures, grouped as postural load and postural variability, were computed from the IMU orientation and position data. 27 posture series are listed in Table 1 (six for the mean absolute angle, six for the mean absolute angular velocity, six for the angle variance, four for the movement speed, and five for the angle distribution).

Table 1. Postural Load and Variability Measures.

| Cate-gory | Measure | Direct-ion | Body part | Description of the measure |
|---|---|---|---|---|
| Postural load | Mean absolute angle (°) | Pitch, roll, yaw | Neck, trunk | The mean of the absolute angle over the analysis window. This measure is obtained separately for the neck and the trunk in each rotational direction. A higher value indicates a greater average deviation from a neutral posture. |
| | Angle distribution (%) | Pitch | Neck | The percentage of task duration during which the neck flexion angle fell within 0°–20° or more than 20°, following a systematic review identifying 20° as the threshold separating low- and high-risk neck postures (Norasi et al., 2022). A higher percentage of time above 20° indicates more time spent in a posture associated with a greater risk of discomfort. |
| | | Pitch | Trunk | The percentage of task duration during which the trunk flexion angle fell within 0°–20°, 20°–60°, or more than 60°, following ISO 11226 (International Organization |

| Category | Measure | Direction | Body part | Description of the measure |
|---|---|---|---|---|
| | | | | for Standardization, 2000) and the trunk posture scoring scheme of McAtamney and Corlett (1993). A higher percentage of time in the higher ranges indicates more time spent in a posture associated with a greater risk of discomfort. |
| Postural variability | Mean absolute angular velocity (°/s) | Pitch, roll, yaw | Neck, trunk | The sum of the absolute frame-to-frame angular change divided by task duration. A higher value indicates more frequent postural change and was interpreted as more favorable, as a sustained, unvarying posture is considered a risk factor independent of postural load. |
| | Angle variance (°²) | Pitch, roll, yaw | Neck, trunk | The variance of the angle time series within the analysis window. A higher value indicates a more varied posture and was interpreted as more favorable. |
| | Movement speed (mm/s) | N/A | Head, trunk, left shoulder, right shoulder | The total translational path length of the segment divided by task duration. A higher value indicates more frequent postural change and was interpreted as more favorable. |

#### 2.4.3. EMG

Muscle activity was characterized by ten metrics, which were grouped into muscular load, variability of muscle activity, and temporal persistence. All metrics were computed from the amplitude envelope of the EMG signal, normalized to the maximum voluntary contraction (MVC) of the same muscle in the same participant and expressed as %MVC; the recording and the signal processing are described below.

Muscular load was described by the mean %MVC over the analysis window and by the 10th, 50th, and 90th percentiles of the amplitude probability distribution function (APDF P10, P50, and P90), which represent the low, typical, and high levels of muscle activity, respectively (Jonsson, 1982).

Variability of muscle activity was described by the standard deviation (SD) of the envelope and by the coefficient of variation (CV), that is, the SD divided by the mean, which expresses the variation relative to the mean activity level; both statistics have been used to quantify motor variability (Srinivasan & Mathiassen, 2012). The range of activity was described by the amplitude ratio ln(P90/P10) and its two components ln(P90/P50) and ln(P50/P10). The difference between the 10th and the 90th percentile of the APDF has been used as an indicator of the variation in muscle activity amplitude (Szeto et al., 2009); the present study used the corresponding logarithmic ratio, which expresses the same range on a scale-invariant basis and can be decomposed into an upper and a lower component.

Temporal persistence was described by the 1-second autocorrelation, defined as corr[EMG(t), EMG(t + 1 s)]; lower values indicate that a given level of muscle activity was sustained for a shorter time. The lag-1 autocorrelation has been used to characterize the temporal structure of motor variability in repetitive manual work, although on kinematic rather than EMG time series (Sedighi & Nussbaum, 2019); its application to seated EMG in the present study is exploratory.

Surface electromyography (EMG) was recorded bilaterally from the upper trapezius and the lateral gastrocnemius using a wireless EMG system (Ultium EMG, Noraxon, USA) at a sampling rate of 4,000 Hz. Electrode placement followed the SENIAM recommendations (Hermens et al., 1999).

Before the tasks, the MVC of each muscle was measured to provide the normalization reference. For the upper trapezius, the experimenter pressed down on both shoulders of the participant while the participant elevated the shoulders against the resistance with maximal effort for five seconds. For the lateral gastrocnemius, the participant raised both heels with maximal effort for five seconds. The two measurements were separated by a one-minute rest, and the whole procedure was repeated three times.

The raw signals were band-pass filtered between 20 and 500 Hz with a fourth-order zero-lag Butterworth filter to retain the frequency components of muscle activity and to attenuate sensor movement, skin-contact variation, and instrumentation noise. The 20 Hz high-pass cut-off follows the recommendation of De Luca et al. (2010) for general use in surface EMG, and the pass band is within the range recommended for surface EMG recordings (Hermens et al., 1999). The filtered signals were converted into amplitude envelopes with a 0.1-second (400-sample) moving root-mean-square (RMS) window advanced one sample at a time, so that the output retained the original 4,000 Hz sampling rate. The envelopes were then normalized to the participant- and channel-specific MVC reference, defined as the maximum of a one-second moving RMS of the identically filtered MVC trial. Isolated missing samples caused by packet loss were linearly interpolated; this occurred in one channel of one session (5.85% of the samples in that channel). The ten metrics were computed for each channel and were then averaged between the left and the right channel to yield one representative value per participant, muscle, chair, and task; the raw signals themselves were not averaged.

The analysis window was 0 to 900 seconds of each task. In task 2, participants rotated 180° every three minutes to report the corrected words and then returned to the front-facing posture, which accounted for approximately 30 to 50 seconds of the 15 minutes. Because the rotation periods required a pattern of whole-body movement that differed substantially from the front-facing work, the rotation periods, defined from the onset of rotation to the return to the front-facing posture on the basis of the IMU signals, were excluded from the analysis. The 1-second autocorrelation was the one exception and was computed over the full 15-minute signal, because this metric is defined on pairs of samples one second apart: excluding the rotation periods would both reduce the number of available pairs and break the temporal structure at the boundaries of the excluded periods. Samples on either side of an excluded period were never joined to form a pair.

### 2.4.4. Task performance

Proofreading performance was evaluated using the framework of Signal Detection Theory (SDT), which separates a participant's ability to discriminate misspelled from correctly spelled words from the participant's response bias in making that judgment (Green & Swets, 1966). Each participant proofread a short-story text file (.txt) in which a subset of words had been deliberately misspelled; the unit of analysis was each space-separated word in the text. For every word, the participant's response (corrected vs. left uncorrected) was cross-tabulated against the word's true status (misspelled vs. correct) to yield four outcome categories: 1) Hit (H): a misspelled word that was correctly detected and corrected; 2) Miss (M): a misspelled word that was not detected and was left uncorrected; 3) False Alarm (FA): a correctly spelled word that was mistakenly identified as an error and unnecessarily corrected; 4) Correct Rejection (CR): a correctly spelled word that was correctly left uncorrected. After calculating these four primary values, ten distinct task performance metrics were calculated (Table 2).

Table 2. Definitions, formulas, and interpretation of performance metrics

| Metric | Definition | Formula | Interpretation |
|---|---|---|---|
| Hit rate (Sensitivity/ Recall) | Proportion of misspelled words correctly identified and corrected | H / (H + M) | Higher is better |
| False alarm rate | Proportion of correctly spelled words unnecessarily corrected | FA / (FA + CR) | Lower is better |
| Precision | Proportion of all correction responses (Hits + False Alarms) that were directed at genuine misspellings | H / (H + FA) | Higher is better |
| Overall accuracy | Proportion of correct responses across all words in the analyzed scope | (H + CR) / (H + M + FA + CR) | Higher is better |
| Reading throughput (words/min) | Processing speed, converted to number of words read and judged per minute | n_words / 15min | Higher indicates faster processing |
| Correct corrections/ min | Number of misspelled words correctly identified and corrected per minute | H / 15min | Higher is better |
| d′ (Sensitivity index) | Ability to discriminate between correctly spelled and misspelled words | Z(Hit Rate) – Z(False Alarm Rate) | Higher is better |

| Metric | Definition | Formula | Interpretation |
|---|---|---|---|
| Criterion c | Degree and direction of response bias in judging a word as misspelled | −0.5 × [Z(Hit Rate) + Z(False Alarm Rate)] | Negative values indicate a more liberal response criterion |
| Specificity | Proportion of correctly spelled words accurately judged as correct without unnecessary modification | CR / (CR + FA) | Higher is better |
| F1-score | Harmonic mean of precision and recall, jointly evaluating detection accuracy and avoidance of unnecessary corrections | 2 × (Precision × Recall) / (Precision + Recall) | Higher is better |

For d′ and criterion c, the hit rate and false alarm rate used in the z-transformation were adjusted with the log-linear correction to avoid undefined values when either rate equaled 0 or 1. Under this correction, a correction applied to the wrong target word was treated as a detected signal and included in the corrected hit-rate numerator.

### 2.5. Statistical analysis

Differences among the three chairs were tested with the Friedman test, and for the metrics with a significant main effect all three pairs of chairs were compared with the Wilcoxon signed-rank test. Differences between the two tasks were tested with the Wilcoxon signed-rank test on values averaged across the three chairs within each participant. All tests were two-tailed (significance level: $p \leq 0.05$), and no correction for multiple comparisons was applied, given the exploratory nature of the posture, muscle activity, and subgroup analyses; results should be interpreted with this in mind. In addition to the whole sample ($n = 20$), male and female subgroups ($n = 10$ each) were analyzed separately on an exploratory basis, except where missing or unusable data required excluding a participant from a specific analysis, in which case the resulting sample size is reported alongside that analysis. Interview responses about chair ranking were grouped into themes by content similarity.

## 3. Results

### 3.1. Subjective ratings

#### 3.1.1. Comparison among chairs

No significant differences among the three chairs in the Friedman test were found for overall satisfaction or seamless support, either for the overall sample or within either sex (all $p > 0.05$). Chair C was most frequently ranked first, selected as the most preferred chair by 9 of 20 participants, followed by Chair A (8 participants) and Chair B (5 participants). Chair C was also most frequently ranked last, selected as the least preferred chair by 10 participants, followed by Chair A (6 participants) and Chair B (5 participants) (Table 3). Two participants (P02, P13) rated Chair A and Chair B as tied for first place, and one participant (P16) rated Chair A and Chair B as tied for last place; each chair in a tie was credited with that rank, so the totals sum to 22 for first place and 21 for last place rather than 20 (Table 4).

Table 3. Individual participant chair-preference rankings.

| Participant | Preference ranking | Rank 1 (Most preferred) | Rank 3 (Least preferred) |
|---|---|---|---|
| P01 | C > B > A | C | A |
| P02 | A = B > C | A, B | C |
| P03 | A > B > C | A | C |
| P04 | A > B > C | A | C |
| P05 | B ≫ A > C | B | C |
| P06 | C > A > B | C | B |
| P07 | C ≫ A > B | C | B |
| P08 | A > B > C | A | C |
| P09 | A > B > C | A | C |
| P10 | C ≫ B > A | C | A |
| P11 | C > A ≫ B | C | B |
| P12 | C > A > B | C | B |
| P13 | A = B > C | A, B | C |
| P14 | B > A > C | B | C |
| P15 | A > B > C | A | C |
| P16 | C ≫ A = B | C | A, B |
| P17 | C > B ≥ A | C | A |
| P18 | B > C > A | B | A |
| P19 | A > B > C | A | C |
| P20 | C > B ≥ A | C | A |

Note. Chair identity was withheld from participants during testing. ">" indicates a clear preference, "≫" a strong preference, "=" or "≥" indicates a tie or near-tie.

Table 4. Number of participants ranking each chair first (most preferred) or last (least preferred).

| Rank | Chair A | Chair B | Chair C |
|---|---|---|---|
| Rank 1 (Most preferred) | 8 | 5 | 9 |
| Rank 3 (Least preferred) | 6 | 5 | 10 |

In the post-study interviews, participants' stated reasons for preferring or not preferring a chair were varied, including the supportive feel or flexibility of the seating surface and backrest, the shape of the seat pan and its fit to the body, familiarity with the material, and the comfort of the headrest. The most common themes are summarized in Table 5.

Table 5. Themes and representative quotes from post-study interviews on chair preference.

| Theme | n | Representative quotes |
|---|---|---|
| Preferred a supportive feel | 5 | [P02] "For lumbar support, Chair A and Chair B were better than Chair C. It feels well-supported."<br>[P05] "In general: Prefer a chair that makes me feel firmly supported during work. It feels like it supports the body well."<br>[P10] "Chair C supported the body the most. It supported the back and lower back well."<br>[P13] "Chair A: The lumbar support felt good, and I could feel the effect clearly when adjusting the functions."<br>[P19] "Chair A: Appropriately supportive, and the height and other settings were satisfactory." |
| Preferred flexible material | 1 | [P06] "Chair C: Overall feels flexible. Feels lighter when seated. I see why they used mesh instead of a cushion." |
| Seat-pan shape and body fit were favorable | 2 | [P07] "Chair C: The seat pan wrapped around the pelvis nicely."<br>[P16] "Chair C: Feels like it fits my body well... the backrest was short, which suited me better since my upper body is short." |
| Preferred a familiar mesh chair | 1 | [P17] "Chair C: Personally similar to the chair I use at home. I like mesh chairs." |
| Headrest comfort and overall use feel were favorable | 1 | [P20] "Chair C: The headrest was the most comfortable... the backrest balance was also good, and I didn't get as tired." |
| Mesh material felt insufficiently supportive | 3 | [P05] "Chair C: Didn't feel supported by the body. It felt like sitting on a sofa while working. My body sank in."<br>[P09] "Chair C: The mesh seat pan material was uncomfortable." |
| | | [P14] "It seemed okay if I sat using the full range of motion... but the mesh material wasn't great either." |
| Did not accommodate body type well | 2 | [P14] "The shallow seat pan depth bothered me, and the mesh material wasn't great either. (*Likely because the participant has long legs.)"<br>[P19] "Chair C: My legs ended up fixed in an uncomfortable position... my legs are long, so the chair felt awkward no matter how I positioned them." |
| Armrests did not stay fixed | 2 | [P04] "Chair C: The armrest kept drifting back and forth, which was slightly uncomfortable — the mechanisms weren't smooth."<br>[P17] "Chair A, B: It would have been nice if the armrests were fixed." |
| Preferred a fixed or less-moving backrest | 2 | [P09] "The backrest recline resistance wasn't adjustable, so it felt like it reclined too far. So I just locked it."<br>[P14] "Chair B: Having the backrest fixed in place with a bar was actually good. I didn't need to keep adjusting my posture, so I found it the most comfortable." |
| Headrest was uncomfortable | 1 | [P06] "Chair A: The headrest was very uncomfortable. My neck was held too straight, which caused discomfort during work." |
| Preferred intuitive controls | 1 | [P04] "Chair A, Chair B: Anyone could operate them intuitively (unlike Chair C)." |

Note. Since some participants did not have opinions whereas some had multiple opinions, the sum of n exceeds 20.

### 3.1.2. Comparison between tasks

For the overall sample, overall satisfaction did not differ significantly between tasks ($p > 0.05$). Seamless support was higher in task 2 than in task 1 (task 1: $M = 4.00$, $SD = 0.68$; task 2: $M = 4.18$, $SD = 0.63$; $p = 0.042$).

Among male participants, both overall satisfaction (task 1: $M = 3.73$, $SD = 0.83$; task 2: $M = 3.97$, $SD = 0.71$; $p = 0.031$) and seamless support (task 1: $M = 3.73$, $SD = 0.75$; task 2: $M = 4.00$, $SD = 0.68$; $p = 0.031$) were higher in task 2. No significant differences between tasks were found among female participants for either measure ($p > 0.05$) (Table 6).

Table 6. Significant differences in subjective ratings between task 1 and task 2

| Metric | Group | Task 1 | | Task 2 | | p-value | Comparison |
|---|---|---|---|---|---|---|---|
| | | Mean | SD | Mean | SD | | |
| Seamless support | All | 4.00 | 0.68 | 4.18 | 0.63 | 0.042 | Task 1 < Task 2 |
| Overall satisfaction | Males | 3.73 | 0.83 | 3.97 | 0.71 | 0.031 | Task 1 < Task 2 |
| Seamless support | Males | 3.73 | 0.75 | 4.00 | 0.68 | 0.031 | Task 1 < Task 2 |

## 3.2. IMU

Chairwise comparisons were conducted in the whole sample ($n = 19$) and separately in the male ($n = 10$) and female ($n = 9$) subgroups; the subgroup analyses were exploratory, as for the EMG metrics. Task effects were assessed only in the whole sample ($n = 19$) using Wilcoxon signed-rank tests after values were averaged across chairs within each participant. To focus on movement-related postural variability, task comparisons were restricted to the four movement-speed and six mean absolute angular-velocity series. These task comparisons were repeated separately within each chair. All tests were two-tailed (significance level: $p \leq .05$).

### 3.2.1. Comparison among chairs

Of the 81 Friedman tests performed per task (27 postural series × three analytical groups), one was significant in task 1 and eight were significant in task 2 (Table 7). In task 1, the mean absolute angle of the trunk in roll differed among chairs in the male subgroup ($p = 0.020$), but none of the pairwise comparisons was significant (all $p \geq 0.083$). No other chair effect was observed in the whole sample or either subgroup. In task 2, two chair effects were observed in the whole sample. Mean absolute angular velocity of the neck in pitch differed among chairs ($p = 0.041$), with lower values for Chair B than for Chair C ($p = 0.012$). Mean absolute angular velocity of the trunk in yaw also differed among chairs (p = 0.010): Chair A was higher than Chair B (p = 0.042), and Chair B was higher than Chair C (p = 0.024). Chair A did not differ from Chair C for either measure.

In the male subgroup, the mean absolute angle of the trunk in yaw differed among chairs (p = 0.025), with Chair B lower than both Chair A and Chair C (both p = 0.032). Mean absolute angular velocity of the neck in pitch also differed among chairs (p = 0.014), with Chair B lower than Chair C (p = 0.014). The remaining pairwise comparisons were not significant. In the female subgroup, four measures differed among chairs. Chair B showed a lower mean absolute angle of the neck in roll than Chair A (p = 0.025), a lower mean absolute angle of the trunk in roll than Chair C (p = 0.006), and a lower mean absolute angular velocity of the trunk in yaw than Chair C (p = 0.042). Chair B also showed a greater percentage of task time with neck flexion below 20° than Chair C (p = 0.033). No other pairwise comparison was significant.

Table 7. Significant results of Friedman tests for posture measures among the chairs

(N = 19, males = 10, females = 9)

| **Task** | **Body part** | **Metric** | **Group** | **Chair A** | | **Chair B** | | **Chair C** | | **p-value** | **Comparison** |
|---|---|---|---|---|---|---|---|---|---|---|---|
| | | | | **Mean** | **SD** | **Mean** | **SD** | **Mean** | **SD** | | |
| Task 1 | Trunk | Mean absolute angle (Roll) | Males | 2.84 | 1.86 | 3.51 | 1.80 | 2.65 | 3.14 | 0.02 | None |
| Task 2 | Neck | Mean absolute angular velocity (Pitch) | All | 4.65 | 1.23 | 4.41 | 1.33 | 4.88 | 1.54 | 0.041 | Chair B < Chair C (p = 0.012) |
| Task 2 | Trunk | Mean absolute angular velocity (Yaw) | All | 2.54 | 3.93 | 2.2 | 3.05 | 1.88 | 0.79 | 0.01 | Chair A > Chair B (p = 0.042); Chair B > Chair C (p = 0.024) |
| Task 2 | Trunk | Mean absolute angle (Yaw) | Males | 6.56 | 5.08 | 4.31 | 3.06 | 7.1 | 5.91 | 0.025 | Chair B < Chair A (p = 0.032); Chair B < Chair C (p = 0.032) |
| Task 2 | Neck | Mean absolute angular velocity (Pitch) | Males | 5.37 | 1.35 | 5.15 | 1.42 | 5.9 | 1.59 | 0.014 | Chair B < Chair C (p = 0.014) |
| Task 2 | Neck | Mean absolute angle (Roll) | Females | 4.57 | 1.92 | 3.29 | 1.12 | 4.63 | 2.26 | 0.025 | Chair B < Chair A (p = 0.025) |
| Task 2 | Trunk | Mean absolute angle (Roll) | Females | 2.91 | 2.15 | 1.64 | 0.55 | 2.78 | 1.92 | 0.014 | Chair B < Chair C (p = 0.006) |
| Task 2 | Trunk | Mean absolute angular velocity (Yaw) | Females | 1.65 | 0.34 | 1.4 | 0.31 | 1.56 | 0.24 | 0.027 | Chair B < Chair C (p = 0.042) |
| Task 2 | Neck | Angle distribution, <20° (Pitch) | Females | 94.42 | 5.61 | 93.78 | 8.02 | 86.8 | 13.48 | 0.013 | Chair B > Chair C (p = 0.033) |

### 3.2.2. Comparison between tasks

Across the whole sample (n = 19), movement speed was higher in task 2 than in task 1 at the head (task 1: M = 19.73, SD = 9.03; task 2: M = 31.26, SD = 13.54), trunk (task 1: M = 14.28, SD = 7.66; task 2: M = 21.23, SD = 12.13), left shoulder (task 1: M = 13.78, SD = 7.81; task 2: M = 21.05, SD = 15.00), and right shoulder (task 1:

M = 13.75, SD = 7.28; task 2: M = 20.75, SD = 12.58; all p < .001; Table 8). Mean absolute angular velocity was also higher in task 2 than in task 1 for the neck in pitch (task 1: M = 2.84, SD = 0.96; task 2: M = 4.64, SD = 1.30), roll (task 1: M = 1.96, SD = 0.73; task 2: M = 3.50, SD = 1.24), and yaw (task 1: M = 5.41, SD = 3.73; task 2: M = 14.57, SD = 6.83). The same pattern was observed for the trunk in pitch (task 1: M = 1.19, SD = 0.67; task 2: M = 1.72, SD = 0.86), roll (task 1: M = 0.74, SD = 0.39; task 2: M = 1.15, SD = 0.60), and yaw (task 1: M = 1.23, SD = 0.86; task 2: M = 2.20, SD = 2.53; all p < .001).

Table 8. Significant Task Effects for Postural Measures in the Whole Sample (n = 19).

| **Body part** | **Metric** | **Task 1** | | **Task 2** | | **p-value** | **Comparison** |
|---|---|---|---|---|---|---|---|
| | | **Mean** | **SD** | **Mean** | **SD** | | |
| Head | Movement speed (mm/s) | 19.73 | 9.03 | 31.26 | 13.54 | < 0.001 | Task 1 < Task 2 |
| Trunk | Movement speed (mm/s) | 14.28 | 7.66 | 21.23 | 12.13 | < 0.001 | Task 1 < Task 2 |
| Left should-er | Movement speed (mm/s) | 13.78 | 7.81 | 21.05 | 15 | < 0.001 | Task 1 < Task 2 |
| Right should-er | Movement speed (mm/s) | 13.75 | 7.28 | 20.75 | 12.58 | < 0.001 | Task 1 < Task 2 |
| Neck | Mean absolute angular velocity, Pitch (°/s) | 2.84 | 0.96 | 4.64 | 1.3 | < 0.001 | Task 1 < Task 2 |
| Neck | Mean absolute angular velocity, Roll (°/s) | 1.96 | 0.73 | 3.5 | 1.24 | < 0.001 | Task 1 < Task 2 |
| Neck | Mean absolute angular velocity, Yaw (°/s) | 5.41 | 3.73 | 14.57 | 6.83 | < 0.001 | Task 1 < Task 2 |
| Trunk | Mean absolute angular velocity, Pitch (°/s) | 1.19 | 0.67 | 1.72 | 0.86 | < 0.001 | Task 1 < Task 2 |
| Trunk | Mean absolute angular velocity, Roll (°/s) | 0.74 | 0.39 | 1.15 | 0.6 | < 0.001 | Task 1 < Task 2 |
| Trunk | Mean absolute angular velocity, Yaw (°/s) | 1.23 | 0.86 | 2.2 | 2.53 | < 0.001 | Task 1 < Task 2 |

### 3.3. Electromyography

#### 3.3.1. Comparison among chairs

Of the twenty metrics examined in each task (ten metrics for each of the two muscles), the Friedman test was significant for one metric in task 1 and for three metrics in task 2 (Table 9). In task 1, a significant main effect of chair was found for the APDF P90 of the upper trapezius (p = 0.043). The post hoc comparison showed a lower value for Chair A than for Chair C (Chair A: M = 5.81, SD = 5.51; Chair C: M = 7.80, SD = 5.81; p = 0.007). Chair B was intermediate (M = 6.25, SD = 5.38), and the remaining two pairs did not differ significantly (Chair A vs. Chair B: p = 0.368; Chair B vs. Chair C: p = 0.058). No significant differences were found for the other trapezius metrics or for any gastrocnemius metric, and no significant differences were found in either subgroup.

In task 2, a significant main effect of chair was found for the amplitude ratio ln(P90/P50) of the upper trapezius (p = 0.043; Chair A: M = 1.478, SD = 0.705; Chair B: M = 1.652, SD = 0.566; Chair C: M = 1.526, SD = 0.478), but none of the three pairwise comparisons reached significance (Chair A vs. Chair B: p = 0.076; Chair B vs. Chair C: p = 0.177; Chair A vs. Chair C: p = 0.701).

In the male subgroup, two gastrocnemius metrics showed a significant main effect of chair in task 2. The SD was lower for Chair B than for Chair C (p = 0.014; Chair B: M = 2.04, SD = 1.90; Chair C: M = 2.40, SD

= 1.97; pairwise p = 0.002), and the 1-second autocorrelation was likewise lower for Chair B than for Chair C (p = 0.027; Chair B: M = 0.250, SD = 0.121; Chair C: M = 0.341, SD = 0.140; pairwise p = 0.014). For both metrics the comparisons involving Chair A were not significant. No significant differences were found in the female subgroup. The subgroups comprised ten participants each, and the sexes were not compared with each other.

Across both tasks, both muscles, and all metrics, no significant difference was found between Chair A and Chair B in any comparison.

Table 9. EMG metrics with a significant difference among the chairs. Values are in %MVC except the CV, the logarithmic ratios, and the 1-s autocorrelation.

| Task | Muscle | Metric | Group | Chair A | | Chair B | | Chair C | | p-value | Comparison |
|---|---|---|---|---|---|---|---|---|---|---|---|
| | | | | Mean | SD | Mean | SD | Mean | SD | | |
| Task 1 | Upper trapezius | APDF P90 | All | 5.81 | 5.51 | 6.25 | 5.38 | 7.80 | 5.81 | 0.043 | Chair A < Chair C (p = 0.007) |
| Task 2 | Upper trapezius | ln(P90/P50) | All | 1.478 | 0.705 | 1.652 | 0.566 | 1.526 | 0.478 | 0.043 | None |
| Task 2 | Lateral gastrocnemius | SD | Males | 2.07 | 1.89 | 2.04 | 1.90 | 2.40 | 1.97 | 0.014 | Chair B < Chair C (p = 0.002) |
| Task 2 | Lateral gastrocnemius | 1-s auto correlation | Males | 0.303 | 0.158 | 0.250 | 0.121 | 0.341 | 0.140 | 0.027 | Chair B < Chair C (p = 0.014) |

### 3.3.2. Comparison between tasks

Values were averaged across the three chairs within each participant and compared between the two tasks (n = 20). For the upper trapezius, eight of the ten metrics were significantly higher in task 2 than in task 1 (Table 10). No significant difference was found for the APDF P10 (p = 0.261) or the CV (p = 0.165).

For the lateral gastrocnemius, no significant differences were found for the mean %MVC (p = 0.571), the APDF P10 (p = 0.388), P50 (p = 0.475), and P90 (p = 0.070), the SD (p = 0.701), or the CV (p = 0.522). The amplitude ratio ln(P90/P10) (task 1: M = 0.731, SD = 0.511; task 2: M = 0.800, SD = 0.562; p = 0.036), its upper component ln(P90/P50) (task 1: M = 0.499, SD = 0.458; task 2: M = 0.574, SD = 0.498; p = 0.009), and the 1-second autocorrelation (task 1: M = 0.266, SD = 0.164; task 2: M = 0.334, SD = 0.137; p = 0.044) were higher in task 2.

The same pattern was observed in the subgroups. In the female subgroup, eight trapezius metrics, all except the APDF P10 and the CV, were higher in task 2; in the male subgroup, eight trapezius metrics, all except the CV and ln(P90/P50), were higher in task 2. For the gastrocnemius, only ln(P90/P10) was significantly higher in task 2 in the female subgroup, and no metric differed significantly in the male subgroup.

Table 10. EMG metrics with a significant difference between the tasks (n = 20). Values are in %MVC except the logarithmic ratios and the 1-s autocorrelation.

| Muscle | Metric | Task 1 | | Task 2 | | p-value | Comparison |
|---|---|---|---|---|---|---|---|
| | | Mean | SD | Mean | SD | | |
| Upper trapezius | Mean | 3.31 | 2.77 | 4.87 | 3.88 | < 0.001 | Task 1 < Task 2 |
| Upper trapezius | APDF P50 | 2.21 | 2.43 | 3.19 | 3.31 | < 0.001 | Task 1 < Task 2 |
| Upper trapezius | APDF P90 | 6.62 | 5.02 | 10.51 | 7.70 | < 0.001 | Task 1 < Task 2 |
| Upper trapezius | SD | 3.74 | 2.30 | 5.44 | 3.32 | < 0.001 | Task 1 < Task 2 |
| Upper trapezius | ln(P90/P10) | 2.027 | 0.509 | 2.564 | 0.548 | < 0.001 | Task 1 < Task 2 |
| Upper trapezius | ln(P90/P50) | 1.294 | 0.376 | 1.552 | 0.517 | 0.030 | Task 1 < Task 2 |
| Upper trapezius | ln(P50/P10) | 0.733 | 0.350 | 1.012 | 0.512 | < 0.001 | Task 1 < Task 2 |
| Upper trapezius | 1-s autocorrelation | 0.467 | 0.127 | 0.544 | 0.103 | < 0.001 | Task 1 < Task 2 |

| Muscle | Metric | Task 1 | | Task 2 | | p-value | Comparison |
|---|---|---|---|---|---|---|---|
| | | Mean | SD | Mean | SD | | |
| Lateral gastrocnemius | ln(P90/P10) | 0.731 | 0.511 | 0.800 | 0.562 | 0.036 | Task 1 < Task 2 |
| Lateral gastrocnemius | ln(P90/P50) | 0.499 | 0.458 | 0.574 | 0.498 | 0.009 | Task 1 < Task 2 |
| Lateral gastrocnemius | 1-s autocorr-elation | 0.266 | 0.164 | 0.334 | 0.137 | 0.044 | Task 1 < Task 2 |

### 3.4. Task performance

One participant (P4, female) had missing data in one condition and was excluded from all task performance analyses, leaving a final sample of 19 participants (10 male, 9 female).

#### 3.4.1. Comparison among chairs

Across the ten task performance metrics, the Friedman test found no significant main effect of chair in the whole sample ($\chi^2 = 0.42–3.40$, all $p \geq 0.18$), the male subgroup ($\chi^2 = 0–4.20$, all $p \geq 0.12$), or the female subgroup ($\chi^2 = 0.29–2.11$, all $p \geq 0.35$). Task performance measures therefore did not distinguish the three chairs at a statistically significant level, and no chair can be concluded to be objectively superior based on these metrics. Some numerical differences were found, but they did not reach statistical significance and should be interpreted only as descriptive, non-significant trends, not as evidence of a statistically superior chair.

#### 3.4.2. Comparison between tasks

In contrast, comparisons between the two tasks yielded several significant differences (Table 11). In the whole sample, reading throughput was significantly higher in Task 1 than in Task 2 (Task 1: M = 120.46, SD = 35.74 words/min; Task 2: M = 93.90, SD = 27.94 words/min; $p < 0.001$), as was correct corrections/min (Task 1: M = 1.39, SD = 0.55; Task 2: M = 1.17, SD = 0.43; $p = 0.002$). Criterion c was also significantly higher in Task 1 (Task 1: M = 1.69, SD = 0.27; Task 2: M = 1.60, SD = 0.22; $p = 0.016$), indicating that participants adopted a more conservative correction criterion under the single-monitor condition of Task 1.

The same pattern held in both subgroups. In the male subgroup, reading throughput (Task 1: M = 132.28, SD = 21.90; Task 2: M = 102.93, SD = 17.12; $p = 0.002$), correct corrections/min (Task 1: M = 1.40, SD = 0.38; Task 2: M = 1.20, SD = 0.25; $p = 0.049$), and criterion c (Task 1: M = 1.77, SD = 0.22; Task 2: M = 1.68, SD = 0.19; $p = 0.049$) were all significantly higher in Task 1. In the female subgroup, reading throughput (Task 1: M = 107.33, SD = 44.34; Task 2: M = 83.86, SD = 34.81; $p = 0.039$) and correct corrections/min (Task 1: M = 1.37, SD = 0.72; Task 2: M = 1.13, SD = 0.58; $p = 0.012$) were significantly higher in Task 1, but the Criterion c difference did not reach significance in this subgroup ($p = .203$). No significant task differences were found for the remaining metrics (Hit rate, false alarm rate, precision, overall accuracy, d′, specificity, F1-score) in the whole sample or in either subgroup.

Table 11. Task performance metrics with a significant difference between the tasks (n = 19)

| Group | Metric | Task 1 | | Task 2 | | p-value | Comparison |
|---|---|---|---|---|---|---|---|
| | | Mean | SD | Mean | SD | | |
| All | Reading throughput (words/min) | 120.46 | 35.74 | 93.90 | 27.94 | < 0.001 | Task 1 > Task 2 |
| All | Correct corrections/ min | 1.39 | 0.55 | 1.17 | 0.43 | 0.002 | Task 1 > Task 2 |
| All | Criterion c | 1.69 | 0.27 | 1.60 | 0.22 | 0.016 | Task 1 > Task 2 (conservative criterion) |
| Males | Reading throughput (words/min) | 132.28 | 21.90 | 102.93 | 17.12 | 0.002 | Task 1 > Task 2 |
| Males | Correct corrections/ min | 1.40 | 0.38 | 1.20 | 0.25 | 0.049 | Task 1 > Task 2 |
| Males | Criterion c | 1.77 | 0.22 | 1.68 | 0.19 | 0.049 | Task 1 > Task 2 (conservative criterion) |

| Group | Metric | Task 1 | | Task 2 | | p-value | Comparison |
|---|---|---|---|---|---|---|---|
| | | Mean | SD | Mean | SD | | |
| Females | Reading throughput (words/min) | 107.33 | 44.34 | 83.86 | 34.81 | 0.039 | Task 1 > Task 2 |
| Females | Correct corrections/ min | 1.37 | 0.72 | 1.13 | 0.58 | 0.012 | Task 1 > Task 2 |

# 4. Discussion

## 4.1. Comparison between Chair A (T90) and Chair C (Aeron)

Across nearly all measures examined in this study, Chair A and Chair C did not differ significantly. Subjective ratings of overall satisfaction and seamless support showed no significant difference between the two chairs, in either the whole sample or either sex subgroup. Task performance likewise did not distinguish the two chairs: none of the performance metrics showed a significant chair effect in the whole sample or in either subgroup, so neither chair can be said to support faster or more accurate proofreading than the other.

Postural measures (IMU) and muscle activity (EMG) showed only a handful of significant differences out of the many series tested, and these did not converge on a consistent advantage for either chair. For posture, a chair effect was found in only 1 of 81 Friedman tests in task 1 and 8 of 81 in task 2; where an omnibus effect was found, the pairwise comparison between Chair A and Chair C specifically was never significant, in either task, in the whole sample, or in either sex subgroup. For muscle activity, Chair A showed a significantly lower upper-trapezius APDF P90 than Chair C in task 1 in the whole sample, suggesting a lower peak muscular load for Chair A during the single-device task; this was the only instance across all EMG and IMU measures in which Chair A showed a statistically significant advantage over Chair C. This difference did not reappear in task 2 or in any other trapezius or gastrocnemius metric, and no other significant Chair A–Chair C EMG difference was found. Taken together, these results do not support a general postural or muscular advantage of either chair over the other; the one significant EMG difference should be interpreted cautiously given the large number of comparisons performed.

Preference ranking presented a more differentiated, and more polarized, pattern than the objective and satisfaction measures. Chair C was both the most frequently ranked first (9 of 20 participants) and the most frequently ranked last (10 of 20 participants), more often than either Chair A (8 first, 6 last) or Chair B (5 first, 5 last). Notably, although Chair A and Chair C were ranked first by a broadly similar number of participants (8 vs. 9), Chair A was ranked last by fewer participants than Chair C (6 vs. 10), indicating that Chair A was less often participants' least preferred option. The post-study interviews suggest that this split did not reduce to a single flexibility dimension. Five participants preferred a supportive feel (three of them in reference to Chair A or Chair B), and three found Chair C's mesh material insufficiently supportive. Participants who favored Chair C, in contrast, gave more varied reasons: the flexible feel of the material (one participant), the shape of the seat pan and its fit to the body (two participants), familiarity with mesh chairs (one participant), and the comfort of the headrest and overall use feel (one participant) (Table 5). Preference therefore appears to depend on individual differences in preferred support feel, body fit, and familiarity, rather than reflecting a chair that is more broadly preferred — differences not captured by the sample-level satisfaction ratings, which showed no significant chair differences.

## 4.2. Comparison between Chair A (T90) and Chair B (T90-fixed)

Comparing Chair A against Chair B, which differs from Chair A only in that its backrest's lateral movement is mechanically locked, isolates the specific contribution of backrest deformability. Subjective ratings (overall satisfaction, seamless support) showed no significant difference between the two chairs in the whole sample or either sex subgroup, and across both tasks, both muscles, and all EMG metrics, no significant difference was found between Chair A and Chair B in any comparison. Task performance likewise did not differ significantly between the two chairs.

Posture (IMU) provided the only whole-sample, albeit tentative, evidence that backrest deformability mattered. In the whole sample in task 2, the more physically demanding of the two tasks, Chair A showed a significantly higher mean absolute trunk angular velocity in yaw than Chair B, suggesting that the trunk moved more frequently in this rotational direction when the backrest was free to move laterally than when it was locked. Because a higher angular velocity reflects more frequent postural change rather than a sustained, unvarying posture (Section 2.4.2), this result may indicate that the standard, dynamic T90 backrest accommodated trunk movement to a greater extent than the fixed-backrest configuration, under the task that most required it. This finding rests on a single pairwise comparison ($p = 0.042$) among many tests conducted without correction for multiple comparisons (Section 4.4) and should therefore be regarded as exploratory until replicated.

Two further differences emerged in the smaller, exploratory sex subgroups, on a different postural measure, the mean absolute angle, which indexes the average deviation from neutral posture rather than the frequency of movement. In the male subgroup in task 2, Chair B showed a lower mean absolute trunk angle in yaw than Chair A, and in the female subgroup, a lower mean absolute neck angle in roll. Because these two findings concern postural load rather than postural variability, and were observed in subgroups roughly half the size of the whole sample, they do not directly speak to the movement-accommodation pattern identified above; rather than contradicting it, they suggest that locking the backrest may somewhat constrain the range of postural excursion in some subgroups, a separate question from how often the trunk moved, on which the whole-sample, better-powered comparison provides the clearer answer. Task 1 showed no significant difference between the two chairs on any posture measure.

This objective pattern does not fully align with participants' own comments about the backrest, however. In the post-study interviews, two participants specifically praised the fixed-backrest condition, reporting that not having to continuously readjust their posture made Chair B feel the most comfortable of the three chairs (Table 5. This suggests that, for at least some sitters, the additional trunk movement the standard T90 backrest affords is not experienced as an unambiguous benefit; postural variability and subjective comfort are not the same construct, and the present data indicate they did not always move together.

Overall, locking the backrest in a fixed position did not measurably change subjective, muscular, or task-performance outcomes relative to the standard dynamic backrest, but it was associated with reduced trunk movement variability specifically in the more demanding, multi-device task, consistent with the intended function of the T90's dynamic backrest. This tentatively suggests that backrest deformability may contribute to how well a chair accommodates trunk movement under demanding, movement-intensive work, even where it does not detectably affect satisfaction, muscular load, or task performance over a single 15-minute session, and even though, as the interview comments above illustrate, not every participant experienced this added movement as preferable.

### 4.3. Comparison between task 1 (single-device task) and task 2 (multi-device task)

Across nearly all measures examined in this study, task 2 (the multi-device task) imposed greater physical and postural demands on participants than task 1 (the single-device task).

Posture (IMU) showed the clearest pattern (Section 3.2.2). Movement speed, averaged across chairs, was higher in task 2 than in task 1 at the head (task 1: $M = 19.73$, $SD = 9.03$; task 2: $M = 31.26$, $SD = 13.54$), trunk (task 1: $M = 14.28$, $SD = 7.66$; task 2: $M = 21.23$, $SD = 12.13$), left shoulder (task 1: $M = 13.78$, $SD = 7.81$; task 2: $M = 21.05$, $SD = 15.00$), and right shoulder (task 1: $M = 13.75$, $SD = 7.28$; task 2: $M = 20.75$, $SD = 12.58$; all $p < .001$; Table 8. Mean absolute angular velocity was likewise higher in task 2 for the neck, in pitch (task 1: $M = 2.84$, $SD = 0.96$; task 2: $M = 4.64$, $SD = 1.30$), roll (task 1: $M = 1.96$, $SD = 0.73$; task 2: $M = 3.50$, $SD = 1.24$), and yaw (task 1: $M = 5.41$, $SD = 3.73$; task 2: $M = 14.57$, $SD = 6.83$), and for the trunk, in pitch (task 1: $M = 1.19$, $SD = 0.67$; task 2: $M = 1.72$, $SD = 0.86$), roll (task 1: $M = 0.74$, $SD = 0.39$; task 2: $M = 1.15$, $SD = 0.60$), and yaw (task 1: $M = 1.23$, $SD = 0.86$; task 2: $M = 2.20$, $SD = 2.53$; all $p < .001$). The multi-device task therefore required substantially more, and faster, whole-body postural adjustment across every body segment and rotational direction examined, most likely driven by the repeated head and trunk reorientation between the two displays.

This heightened physical activity extended to muscle activity: eight of the ten upper-trapezius metrics were significantly higher in task 2 than in task 1, indicating both greater muscular load and greater variability in muscle activity during the multi-device task (Section 3.3.2). Subjectively, participants also rated seamless support as higher in task 2 than in task 1, and, among male participants, overall satisfaction was likewise higher in task 2 (Section 3.1.2). Task performance, in contrast, was poorer in task 2: reading throughput and correct corrections per minute were both lower, and participants adopted a more liberal response criterion, than in task 1 (Section 3.4.2).

Taken together, task 2 demanded more frequent and larger postural change and higher, more variable muscular effort than task 1, alongside weaker task performance and, notably, higher subjective ratings of seamless support. This combination indicates that multi-device work places substantially greater postural and muscular demands on the sitter than conventional single-screen work, while also relying more heavily on the chair's capacity to accommodate frequent, large-amplitude movement, reflected in the higher seamless-support ratings under task 2 despite the added physical load. As office work continues to shift toward multi-device configurations, these results underscore the growing importance of chairs that can accommodate this expanded range of movement, supporting continued development of dynamic-sitting chairs designed specifically around the physical demands of multi-device work.

### 4.4. Limitations and future research directions

The Chair A–Chair B comparison identified one exploratory effect of backrest deformability: greater trunk movement in yaw during the more demanding, multi-device task. This effect, however, was limited to a single rotational direction in a single task, and did not extend to subjective ratings, muscular load, or task performance over the single session tested here. This narrow scope should not be read as evidence that backrest design is unimportant; rather, the present design compared only two points along what is plausibly a continuum: a backrest that moves freely with the trunk, and one that is fully locked. Future work could vary the degree of backrest stiffness or deformability across more than two levels, and examine a broader set of rotational directions and task types, to test whether an intermediate configuration yields a more favorable balance between postural support and postural variability than either endpoint tested here, and whether the movement-accommodation benefit observed here for trunk yaw generalizes to other planes of motion.

More broadly, the present findings are based on a single 15-minute task per condition, in a laboratory setting, with a sample of 20 participants. Whether the same pattern of largely null chair effects would hold over longer, repeated, or more varied real-world use, and across a wider range of body types and working styles, remains to be tested.

Relatedly, because the Aeron (Chair C) did not originally include a headrest, a widely used third-party headrest of mesh-only construction (also sold together with the Aeron by authorized dealers) was added for this study to allow comparison of headrest-related functions across all three chairs. This headrest differs in construction from that of the T90 (Chair A), which combines double raschel mesh and sponge. Interview responses suggested that the headrest was a salient feature to participants (see Section 3.1), so this addition, while necessary for a controlled comparison, means the Chair C tested here includes a component that is not part of the base Aeron; caution is warranted in generalizing the present findings to the Aeron without this headrest.

The present analyses also involved a large number of statistical tests, comparisons among three chairs and between two tasks across many subjective, postural, muscular, and performance measures, evaluated in the whole sample and, on an exploratory basis, within male and female subgroups, without correction for multiple comparisons. Because a large number of statistical tests were conducted without correction for multiple comparisons, some of the significant differences reported, particularly isolated findings within the smaller male and female subgroups, may represent false positives rather than true effects. Replication with a pre-registered, more targeted set of comparisons would help confirm which effects are robust.

Finally, task 2 operationalized multi-device work as a specific combination of two monitors, a smartphone, and periodic 180° verbal-report rotations; this design isolated a clear, measurable source of additional movement, but it is only one instantiation of multi-device work. Other common forms, such as frequent video calls, cross-application typing, or unscheduled notifications, may impose different postural and muscular demands, and future research should examine whether the present pattern, greater movement and muscular activity, but poorer task performance, generalizes across other multi-device configurations.

## 5. Conclusion

This study compared two contemporary high-end dynamic-sitting chairs (T90, Aeron), isolated the specific contribution of backrest deformability (T90 vs. T90-fixed), and tested whether multi-device work imposes different physical demands than conventional single-device work. Comparing the T90 against the Aeron, subjective ratings, task performance, posture, and muscle activity showed only isolated, direction-inconsistent differences between the two chairs. Isolating backrest deformability specifically, locking the T90's backrest produced no measurable difference in subjective ratings, muscle activity, or task performance, but it was associated with reduced trunk movement (mean absolute angular velocity in yaw) during the more demanding, multi-device task in an exploratory analysis without correction for multiple comparisons, which may indicate that the standard T90's dynamic backrest helped accommodate trunk movement when task demands were high. Chair preference rankings were a further exception to the largely null pattern between the T90 and Aeron: they were polarized rather than uniform, with the same chair often ranked both most and least preferred, suggesting that fit between a chair and an individual sitter's preferences in support feel, body fit, and familiarity matter more than any sample-level advantage of one chair over another.

In contrast, the two tasks differed consistently and substantially, more so than either chair comparison did. The multi-device task required faster and larger postural movement across every body segment examined, elicited greater and more variable muscular load, and was associated with poorer proofreading performance than the single-device task, even though it was also rated as more comfortable in terms of seamless support. This pattern indicates that the demands of the work itself were the dominant driver of the physical differences observed in this study: they affected far more measures, and more consistently, than either the choice between the T90 and the Aeron or the presence of backrest deformability, whose measurable contribution was limited to a single postural dimension in the more demanding task.

Taken together, these findings suggest that as office work continues to shift toward multi-device configurations, the physical demands placed on the sitter grow accordingly, underscoring the need for dynamic-sitting chair designs developed specifically around this expanded range of movement, rather than around the single-screen work that has traditionally guided dynamic-sitting chair design.

## Author statement

Yejin Kang: Writing – Original Draft Preparation, Methodology, Formal Analysis, Visualization, Resources, Software, Review & Editing

Dusan Baek: Writing – Original Draft Preparation, Methodology, Formal Analysis

Dongwoo Park: Writing – Original Draft Preparation, Methodology, Formal Analysis

Kitae Hwang: Writing – Original Draft Preparation, Methodology, Formal Analysis

Woojin Park: Writing – Original Draft Preparation, Review & Editing, Supervision

## Declaration of generative AI and AI-assisted technologies in the manuscript preparation process

During the preparation of this work the author(s) used Claude (Sonnet 5, Anthropic) solely for language editing to improve clarity, grammar, and readability. After using this tool/service, the author(s) reviewed and edited the content as needed and take(s) full responsibility for the content of the article.

## Data statement

The datasets generated and analyzed during the current study are not publicly available due to institutional ethics restrictions and participant confidentiality agreements but are available from the corresponding author on reasonable request.

## Funding

This study was supported by Sidiz.

## Acknowledgments

None